%% file: colm2025_conference.tex
\documentclass{article} 
\usepackage[preprint]{colm2025_conference}

\usepackage{microtype}
\usepackage{hyperref}
\usepackage{url}
\usepackage{booktabs}
\usepackage{graphicx}
\usepackage{array} 
\usepackage{longtable}
\usepackage{colortbl}
\usepackage{multicol}
\usepackage{amsmath}
\usepackage{lineno}
\usepackage{wrapfig}
\usepackage{multirow}
\usepackage{bm}
\usepackage{soul}

\definecolor{darkblue}{rgb}{0, 0, 0.5}
\hypersetup{colorlinks=true, citecolor=darkblue, linkcolor=darkblue, urlcolor=darkblue}

\title{Scaling Auditory Cognition via Test-Time Compute in Audio Language Models}

\author{Ting Dang$^{\spadesuit}$, Yan Gao$^{\diamond}$, and Hong Jia$^{\spadesuit}$\\[0.7ex]
$^{\spadesuit}$School of Computing and Information Systems,
The University of Melbourne, Australia\\[0.7ex]
$^{\diamond}$Department of Computer Science and Technology,
University of Cambridge, UK\\[0.7ex]
\texttt{\{ting.dang,hong.jia\}@unimelb.edu.au}, \texttt{yg381@cam.ac.uk}
}

\begin{document}

\ifcolmsubmission
\linenumbers
\fi

\maketitle
\begin{abstract}
Large language models (LLMs) have shown exceptional versatility in natural language processing, prompting recent efforts to extend their multimodal capabilities to speech processing through the development of audio large language models (Audio LLMs). While Audio LLMs excel in tasks such as speech recognition and synthesis, it remains unclear how they perform when faced with the auditory cognitive challenges posed by real-world environments, such as audio comprehension and listening recall, particularly in the presence of background noise or overlapping speech. Unlike text-based LLMs, which have access to vast amounts of text data for pre-training, retraining Audio LLMs with diverse auditory cognitive scenes is difficult due to the limited datasets that simulate real-world auditory cognitive scenarios and the challenge of acquiring auditory cognitive labels for training. While test-time compute (TTC) methods have been shown to enhance the capabilities of text-based LLMs during inference, a key challenge lies in designing these TTC methods to improve the auditory capabilities of Audio LLMs. This study aims to address these two research gaps by: i) exploring the auditory cognitive capabilities of Audio LLMs, and ii) enhancing their capabilities using TTC approaches. We have investigated five different Audio LLMs for auditory cognition using a \textit{self-collected} database and have proposed five TTC approaches to enhance auditory cognitive capabilities during inference. Our findings reveal that Audio LLMs performance decreases in more challenging auditory cognitive tasks. The proposed TTC approaches significantly enhance cognitive auditory capabilities, advancing the development of more adaptable and resilient Audio LLMs for practical applications such as assistive listening devices, voice-based AI assistants, and communication technologies. 
\end{abstract}

\input{sections/01-intro}
\input{sections/02-relatedwork}
\input{sections/03-data}

\input{sections/04-setup}

\input{sections/05-results}
\input{sections/06-conclusion}


\section*{Acknowledgments}
We acknowledge Nokia Bell Labs, Cambridge, UK for their collaboration in data collection and extend our gratitude to everyone who participated in the data collection process.

\section*{Ethics Statement}
Data collection for this study was conducted in accordance with approved ethical guidelines. Ethical approval for the data collection process was obtained prior to its commencement. Informed consent was acquired from all participants before data collection began. The collected data is securely stored on a server, with all personally identifiable information removed, and will not be shared outside of the ethical guidelines.

\bibliography{colm2025_conference}
\bibliographystyle{colm2025_conference}

\input{sections/07-appendix}

\end{document}

%% file: sections/01-intro.tex
\vspace{-3pt}
\section{Introduction}
\vspace{-1pt}
Large language models (LLMs) have achieved remarkable success as versatile natural language processing systems, demonstrating strong performance across a wide range of tasks such as machine translation~\citep{xu2024contrastive}, question answering~\citep{zhuang2023toolqa, robinson2022leveraging}, and summarization~\citep{laban2023summedits, zhang2024comprehensive}. Building on this progress, recent research has explored extending LLMs to speech processing by integrating acoustic information into their architectures. This has led to the development of audio large language models (Audio LLMs), which can simultaneously process both spoken and written language. Through cross-modal alignment techniques, these models are increasingly capable of handling auditory tasks that require a unified understanding of both linguistic and paralinguistic information, such as emotion recognition~\citep{bellver2024multimodal}, speech quality evaluation~\citep{wang2025enabling}, and music understanding~\citep{verma2025whisper}.

Existing research on audio LLMs primarily focuses on their ability to process and align spoken and textual content, with efforts underway to expand their capabilities to include not only speech but also acoustic scenes and music~\citep{gong2023joint, li2024audio}. 
However, most existing studies were conducted on curated datasets or corpora (e.g., clean speech), which presents a significant limitation when applying the proposed LLMs to real-world auditory environments characterized by background noise, reverberation, and overlapping speech. Unlike text-based benchmarks, often built on large-scale annotated corpora, auditory cognition in complex settings relies on human adaptive mechanisms that are inherently difficult to quantify or label. The scarcity of datasets that faithfully replicate diverse real-world auditory scenarios has led to a limited number of studies addressing these challenging yet realistic conditions. Moreover, even fundamental questions regarding how audio LLMs exhibit cognitive adaptation across varying real-world listening environments remain largely unexplored.
Human listeners exhibit remarkable adaptability in complex situations, effortlessly extracting meaningful information despite distortions like the "cocktail party problem"~\citep{haykin2005cocktail}, supported by enhanced cognitive capabilities and working memory.
A natural question arises: do audio large LLMs possess capabilities comparable to those of humans? More specifically, \textbf{can audio LLMs develop human-like cognitive strategies to maintain comprehension in adverse auditory environments?}

Enhancing these models to improve cognitive auditory capabilities under challenging conditions remains an important open question. In particular, a key challenge lies in effectively adapting pre-trained audio LLMs, originally trained on curated datasets, to real-world auditory environments with minimal additional effort.
Recent studies have showed a potential solution, test-time compute (TTC), wherein dynamic inference strategies enable models to "think longer" on more difficult problems, thereby significantly improving inference performance without requiring additional fine-tuning~\citep{snell2024scaling, bi2024forest}.
Despite its effectiveness, applying TTC to real-world auditory environments remains largely unexplored. This presents a fundamental challenge: \textbf{How can the auditory cognitive robustness of audio LLMs be further enhanced during inference time?}
Addressing these two gaps is crucial for advancing the next generation of AI-driven auditory systems, enabling them to function reliably in real-world applications such as assistive listening devices, voice-based AI assistants, and communication technologies.

This paper aims to bridge this gap by investigating the inherent cognitive capabilities of audio LLMs in auditory processing across varying contexts, comparing their performance to human perception using a \textit{self-collected} dataset, and exploring TTC methods to enhance auditory cognitive tasks.
To address the first question, we explored various audio LLMs in processing different auditory cognitive tasks, ranging from simpler tasks such as audio event recognition to more complex auditory cognitive tasks, including memorizing, recalling, and computing digits spoken in a sequence with overlapped speech. Additionally, to address the second gap, we designed five different TTC strategies to enhance auditory cognitive capabilities. The results on five common audio LLMs revealed that all models performed below human perception, while GPT-4 demonstrated surprisingly strong performance in recognizing overlapped speech. 
The performance of the five proposed TTC approaches varies; however, all demonstrated improvements, with gains ranging from 9\% to 150\%.
The optimal strategy was found to be highly dependent on both the model structure and the complexity of the task. The contributions of this paper are summarized as follows: 
\begin{itemize} 
\item This is the \textit{first-study} to investigate the auditory cognitive capabilities of audio LLMs in real-world environments, demonstrating the inferior performance of current audio LLMs on certain auditory tasks. 
\item This work innovatively proposed and evaluated five TTC strategies aimed at enhancing the performance of audio LLMs on auditory cognitive tasks. It represents the \textit{first-attempt} to enhance audio LLM capabilities during inference time.
\item The TTC experiments, conducted using five different audio LLMs across various tasks, demonstrated significant improvements, highlighting their potential to enhance auditory cognitive processing.
\end{itemize}

%% file: sections/02-relatedwork.tex
\vspace{-5pt}
\section{Related Work}
\vspace{-5pt}
\subsection{Auditory Processing and Cognition}
Human auditory processing is incredibly flexible and can adapt to a wide range of complex scenarios, effectively managing cognitive load to ensure accurate speech comprehension. This adaptability is exemplified by the well-known ``cocktail party effect''~\citep{haykin2005cocktail}, where individuals can focus on a single conversation amid background noise, as well as in various everyday settings such as busy streets, crowded restaurants, and bustling office environments. 
This remarkable ability is driven by advanced neural mechanisms  encompassing adaptive auditory cognition, which enables selective attention, auditory scene analysis, and contextual understanding~\citep{beaman2021auditory}.

Existing research on auditory cognition mainly focuses on psychological analyses to understand the underlying biomechanics or psychological processes~\citep{baldwin2012auditory, anderson2013dynamic}. While recent studies have begun investigating computational modeling including the use of LLMs for cognitive science, they primarily focus on general cognitive functions or visual sensory aspects, such as working memory with visual inputs, and visual-spatial reasoning that analyzes and synthesizes abstract visual stimuli. There is less emphasis on understanding the auditory cognition of audio LLMs~\citep{mcduff2024cognitive, humphreys2021visual}. 
A recent study investigates the auditory attention using audio LLMs, by integrating intracranial electroencephalography (iEEG) recordings with audio LLMs to identify which speaker a listener is focusing on and to refine responses accordingly~\citep{jiang2025aad}. However, this study is specific to a particular listener attention task and requires fine-tuning the audio LLMs with additional iEEG modalities. There still remains a lack of research exploring the potential of audio LLMs for auditory cognition within real-world listening environments, as well as effective strategies to enhance the models' capabilities during inference with minimal effort.


\vspace{-5pt}
\subsection{Audio LLMs}
LLMs' strong capabilities in natural language processing have accelerated the development of audio LLMs that process both the paralinguistic and linguistic information, leading to the emergence of various audio LLMs that enhance capabilities in speech recognition, sound generation, and auditory processing. Commonly used open-source audio LLMs are generally developed based on pre-trained LLaMA, such as LTU-AS~\citep{gong2023joint} and SALMONN~\citep{tangsalmonn}, or transformer structures such as Qwen2-audio~\citep{chu2024qwen2} and Audio-Flamingo~\citep{kong2024audio}. These models generally employ a pre-trained audio encoder such as Whisper~\citep{radford2023robust} that converts raw audio data into a series of audio embeddings, which are then aligned with the text embeddings learned from tokenizers. This alignment allows the embeddings to be processed by a transformer-based architecture, enabling the model to understand audio patterns effectively. 

There are also closed-source audio models, such as GPT-4o~\citep{hurst2024gpt} and Gemini-1.5 pro~\citep{team2024gemini}, which have also demonstrated strong audio scene understanding capabilities. These models are typically trained on large and diverse datasets, allowing them to effectively process auditory content. However, there is still a paucity of studies on how theses audio LLMs process auditory cognitive capabilities in comparison to human perception. 

\vspace{-5pt}
\subsection{Test-time Compute (TTC)}
TTC encompasses a diverse array of strategies aimed at optimizing model inference. These strategies can be categorized into two primary streams: Chain-of-Thought (CoT) prompting and the search against verifiers approach which samples multiple responses and select the optimal response through a verification process.
\vspace{-8pt}
\paragraph{Chain-of-Thought (CoT) Prompting. } CoT prompting has emerged as a promising approach for enhancing LLMs performance and reasoning capabilities during the inference~\citep{wei2022chain}. By breaking down complex tasks into a series of intermediate steps or thoughts, CoT reasoning enables models to tackle challenges in a more structured and transparent manner, particularly effective in tasks that require logical reasoning or multi-step problem-solving~\citep{fu2022complexity}. 
The application of CoT prompting is not limited to text-based tasks. Recent advancements have explored its potential in multi-modal settings, including in the integration of auditory information such as 
the integration of reasoning techniques in audio-visual scene analysis~\citep{shu2023audio}. 
However, these studies primarily focus on clean speech, and there remains a lack of research on whether CoT reasoning is effective in handling complex auditory scenarios, such as overlapping speech. Additionally, how to design CoT prompting for more challenging and realistic audio environments is still underexplored.
\vspace{-8pt}
\paragraph{Search against Verifiers.} This approach typically involves two key steps. The first step generates multiple candidate outputs instead of being limited to a single output. The second step employs a verifier (a reward model), to score the generated outputs in order to identify the optimal output. This method does not require the re-training or fine-tuning of LLMs.

A verifier can range from a hard-coded heuristic to a learned reward model. For the hard-coded heuristic, self-consistency decoding, namely a majority voting mechanism, is commonly employed, where the most frequently generated answer will be selected as the final answer~\citep{wang2022self}. 
Alternatively, a learnt reward model can be employed, which plays a crucial role in the ranking and evaluation of multiple outputs generated by the LLMs~\citep{zhang2024generative}. This reward model assesses the reliability and accuracy of the LLM's multiple outputs by verifying whether they meet specific criteria, such as logical validity, absence of contradictions, and adherence to task-specific requirements. This can be further extended with beam search, where the intermediate steps for generating each candidate output will also be sampled and scored~\citep{yao2023tree}. Typically, this verifier is another task-specific LLM, fine-tuned for scoring the generated outputs. These scores are then used to rank the outputs, with the highest-ranked one being selected.

However, while these approaches have been validated in text-based natural language processing tasks, their scalability to audio cognition tasks, particularly in varied acoustic environments, remains uncertain. A major challenge is the absence of suitable reward models for audio tasks. Existing reward models are primarily designed for text-based applications, where they generate scores based on the original text input and the model’s text output~\citep{wang2024interpretable}. These models are not applicable to audio tasks, which require processing the audio inputs for scoring. Consequently, the development of appropriate reward models for audio LLMs remains an open challenge.

%% file: sections/03-data.tex
\vspace{-5pt}
\section{Methods}

\vspace{-8pt}
\subsection{Data and Design}
\begin{figure}
    \centering
    \includegraphics[width=1\linewidth]{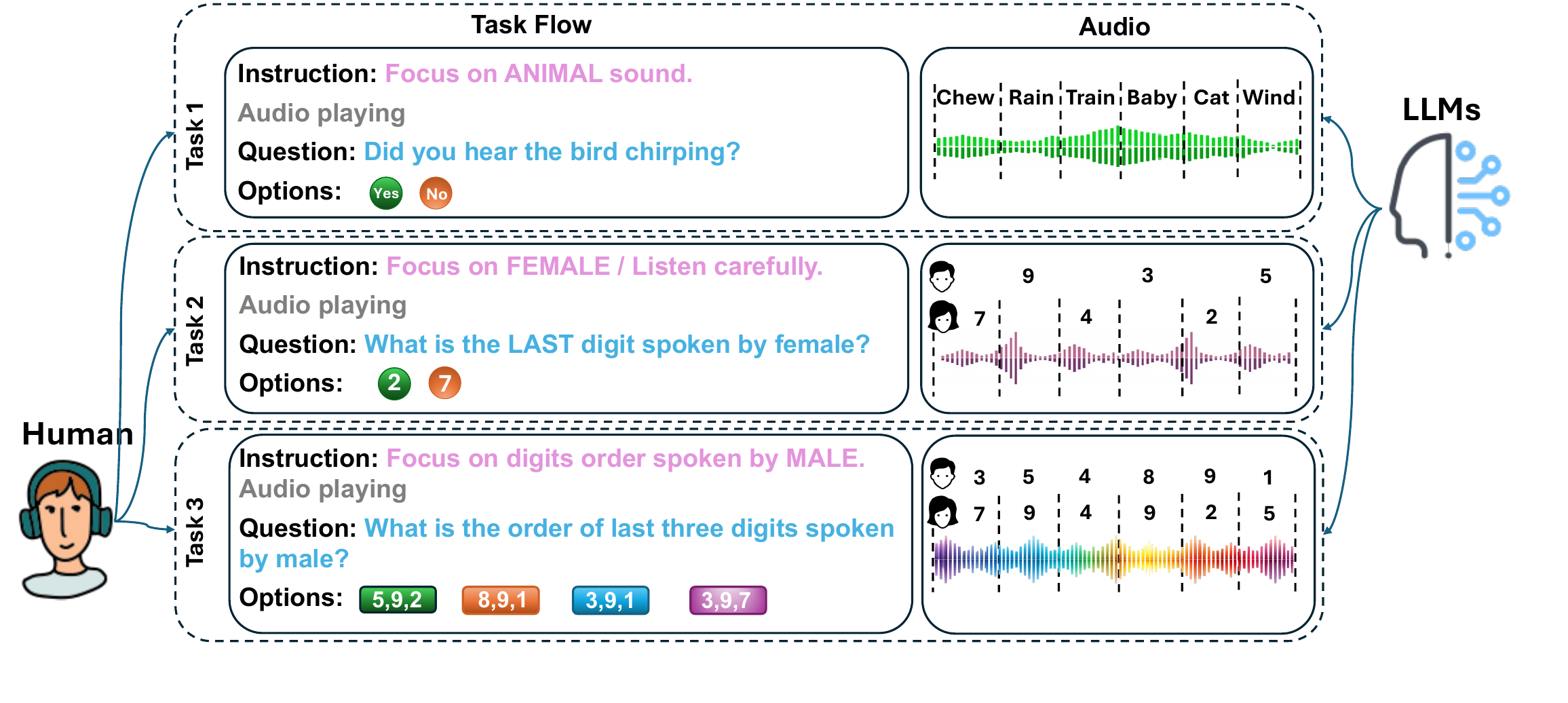}
    \vspace{-20pt}
    \caption{Data collection and experimental design. Three tasks with increasing difficulty in auditory processing were designed, and both humans and LLMs were prompted to answer the same questions after the audio played. }
    \vspace{-8pt}
    \label{fig:1}
\end{figure}

\paragraph{Data. }To simulate a real-world auditory context and analyze LLMs' auditory processing capabilities, we self-collected a database that simulates real-world scenarios. Specifically, we instructed participants to focus on the sound and answer the questions after the audio was played, as shown in Figure~\ref{fig:1}.

Task 1 introduces a low-difficulty auditory scene, where natural sounds devoid of linguistic content are played, such as birds chirping or dogs barking. This task is designed to require minimal cognitive resources. A pre-audio prompt encourages participants to focus on the specific sound (e.g., "Please focus on an animal sound") and a post-audio question (e.g., "Did you hear a bird chirping?") ensures active engagement and helps validate their audio comprehension.

Task 2 increases the difficulty by introducing language comprehension, requiring moderate cognitive resources. Participants hear a person (male or female) reading digits mixed with slight background noise mimicking real-world environments. As with previous tasks, a pre-audio prompt directs participants to focus on digits spoken by a specific gender, and a post-audio question asks them to recall specific details (e.g., "What is the last digit spoken by the female?").

Task 3 presents a high-difficulty auditory scene, imposing a significant cognitive load by simultaneously presenting two spoken digits from male and female voices, amidst background noise. Participants must process and differentiate between the digits spoken by each gender. A pre-audio prompt instructs participants to focus on one gender or both, and a post-audio question asks them to recall the digits spoken by the targeted gender, as well as the sequence of the numbers (e.g., "What is the sum of the last two digits spoken by the male?"). This task is designed to engage a high level of cognitive resources, including the summation of digits, recalling digits that have not been spoken, aimed at testing the limits of auditory processing and attention.

We collected data from 10 participants, including 6 males and 4 females, aged 20 to 55, with an average age of 32. The diverse age range and relatively balanced gender distribution ensure that the study's findings are representative across different demographic groups. All 10 participants provided consent for EEG data collection with the 16-channel wearable EEG headset (g.Nautilus RESEARCH)~\citep{gNautilusRESEARCH} during the study, which was used for task validation. Participants were instructed to maintain a stationary sitting posture to minimize any extraneous bodily movements that could introduce sound artifacts, ensuring the integrity of the data. Our preliminary analysis of the EEG data has validated the effectiveness of the task design in eliciting varying cognitive load, with increasing EEG energy corresponding to increased cognitive load, consistent with existing findings~\citep{williamson2016using}. 

\paragraph{Design. } 
To accurately simulate the context that humans use for auditory processing during data collection, we incorporated both pre-audio and post-audio questions in the prompt for LLMs, along with the audio event, for processing by the audio LLMs. The audio LLMs take the audio input along with a prompt that consists of i) the pre-audio instruction, ii) a post-audio question, and iii) the available options. The model is then prompted to select the best option from the choices provided. This ensures that the information available to humans for answering the questions is the same as what the LLMs receive, thereby maintaining a fair comparison. The prompt used can be referred to Appendix~\ref{app:llmp}.
\subsection{Test-time Compute}
\subsubsection{CoT Prompting}
We additionally include the CoT prompting for three tasks given the context information. These CoT promptings provide additional information on the background knowledge, such as whether one speaker is talking or two speakers are speaking simultaneously. They direct the LLMs to think step-by-step by identifying the gender and digits before arriving at the final answer. 
The details of the CoT promoting can be referred to in Appendix Table~\ref{tab:cot_tasks}.

\subsubsection{Search against Verifiers}
\begin{figure}[t]
    \centering
    \includegraphics[width=1.0\linewidth]{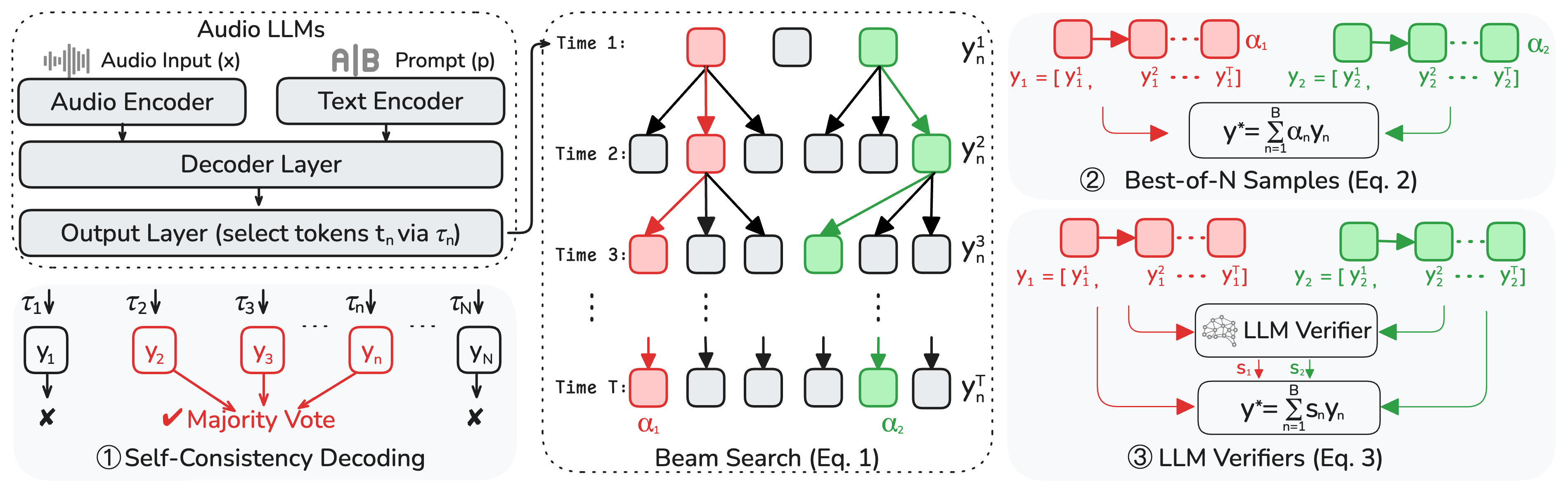}
    \vspace{-2em}
    \caption{Search against verifier approaches: i) Self-consistency decoding, which applies a majority vote to the $N$ multiple outputs; ii) Best-of-N sampling with beam search, where the selected beams (outputs) are weighted by their log-likelihood; and iii) LLM verifier, which employs a stronger LLM as the reward model to score the multiple outputs, ranking or weighting them to optimize the final output.}
    \vspace{-10pt}
    \label{fig:method}
\end{figure}

\paragraph{Self-Consistency Decoding.}  
Self-consistency decoding, also known as the majority voting approach, enhances the reliability of model-generated outputs by aggregating multiple responses. As shown in Figure~\ref{fig:method} (left), given an audio input \(\bm{x}\) and a corresponding prompt \(p\), the LLM backbone generates \(N\) outputs, denoted as \(\bm y_n\), where \(n \in [2, N]\). These outputs can be obtained by varying the temperature values $\tau_n$ during sampling.
A majority voting mechanism is then applied to determine the most frequently selected answer \(\bm y^*\).



\paragraph{Best-of-N Samples with Beam Search.}  
Similarly, the Best-of-N samples approach also generates multiple responses \(\bm y_n\), for a given audio input \(\bm{x}\) and a corresponding prompt \(p\). It is generally integrated with beam search to further explore intermediate steps in diverse decoding paths. In beam search, the model maintains a set of the top \(B\) candidate sequences at each decoding step. 
This allows the model to consider multiple high-likelihood outputs instead of committing to a single most probable sequence, thereby enhancing output diversity and accuracy.  

As shown in Figure~\ref{fig:method} (beam search), at each decoding step \(t\), the model generates possible token sequences \(\bm y_n = (y_n^1, y_n^2, \dots, y_n^T), t \in [1,T] \), where each \( y_n^t \) represents a token at step $t$ in the output sequence. The top \(B\) sequences, represented as $\mathcal{B}_T$, are selected based on their cumulative log probabilities:  
\vspace{-5pt}
\begin{equation}
\mathcal{B}_T = \arg\max_{\mathcal{B}'} \sum_{t=1}^{T} \log P(y_n^t \mid y_n^{1:t-1}, \bm{x}, p), \quad \text{where } |\mathcal{B}'| = B.
\end{equation}  
Once decoding is complete at step \(T\), a reward model is generally employed to score the top \( B \) sequences. This can be performed using either an output reward model (ORM), which assesses only the final output, or a process reward model (PRM), which also scores intermediate steps along the decoding paths~\citep{yao2023tree, zhang2024generative}. However, since ORM and PRM models have not yet been developed for audio LLMs for auditory cognitive tasks to score the outputs effectively, we propose using the cumulative log-likelihood as a scoring mechanism. 

Specifically, 
we employ a weighted combination of all the beam-searched outputs $\bm{y}_n$, with the weight $\alpha_n$ as the corresponding log-likelihood scores as:
\begin{equation}
\bm{y}^* = \sum_{n=1}^{B} \alpha_n \bm{y}_n = \sum_{n=1}^{B} \left(\sum_{t=1}^{T} \log P(y_n^t \mid y_n^{1:t-1}, \bm{x}, p)\right) \bm{y}_n.
\end{equation}
This aims to leverage multiple outputs to optimize the final result, which can be particularly beneficial in challenging auditory cognitive scenarios where relying solely on the top output from audio LLMs may be less reliable.

\paragraph{LLM Verifier.} 
Alternatively, we introduce another LLM verifier as the reward model, to score the multiple outputs generated by the beam search (Figure~\ref{fig:method}). This LLM verifier is chosen as the one with superior audio comprehension capabilities, aiming to leverage a stronger LLM to evaluate the responses of a weaker LLM, thereby enhancing the predictions.
In this case, the LLM verifier takes a prompt that describes the evaluation task along with scoring criteria, as well as the audio input, and outputs a score. The prompt used for the LLM verifier is detailed in Appendix~\ref{app:llm}.

The response with the highest LLM verifier score is chosen as the final output, or a weighted sum of responses \(\bm{y}_n\) is calculated using weights \(s_n\) based on the scores \(S(\bm{y}_n)\) from the LLM verifier as:
\vspace{-5pt}
\begin{equation}
    \bm{y}^* = \arg\max_{\bm{y}_n} S(\bm{y}_n) \quad \text{or} \quad \bm{y}^* = \sum_{n=1}^{N} s_n \bm{y}_n
\end{equation}
\vspace{-10pt}

%% file: sections/04-setup.tex
\vspace{-10pt}
\section{Experimental setup}
\vspace{-5pt}
\subsection{Models and implementation details}
Five different audio LLMs were tested, including Qwen2-Audio~\citep{chu2024qwen2}, Audio-Flamingo 2~\citep{ghosh2025audio}, Gemini-2.0-Flash~\citep{google2023geminiflash}, Gemini-1.5-Pro~\citep{team2024gemini}, and GPT-4o~\citep{hurst2024gpt}. Both Qwen2-Audio and Audio-Flamingo 2 are based on the open-source QwenLM (32 decoder layers)~\citep{bai2023qwen} and Flamingo~\citep{kong2024audio} architecture. Further details on the model structures can be referred to Appendix~\ref{app:setup}, and the implementation details can be referred to Appendix~\ref{app:para}\footnote{The code will be available upon acceptance.}.
\vspace{-5pt}
\subsection{TTC methods}
Five TTC methods were used, including CoT and four other approaches based on search against verifiers:

\begin{itemize}
    \item Temperature-based majority vote (Majority): For a single audio input, $N$ outputs are generated using varying temperatures, and majority voting is used to generate the final answer.
    \item Beam search-based weighted log-likelihood (BS-W): The $|B|$ sequences are weighted based on their log-likelihoods for each sequence.
    \item LLM verifier with highest score (LLM-Top 1): LLM verifier scores the $N$ generated responses, and the one with the highest score is selected.
    \item LLM verifier-based weighted scores (LLM-W): The final answer is selected as the weighted sum of the $N$ responses, with the weights corresponding to the scores evaluated by the LLM verifier.
\end{itemize}
\vspace{-5pt}

%% file: sections/05-results.tex
\vspace{-1pt}
\section{Results}
\vspace{-5pt}
\subsection{Comparison to human perceptions}
\begin{figure}[t]
    \vspace{-2pt}
    \centering
    \includegraphics[width=1\linewidth]{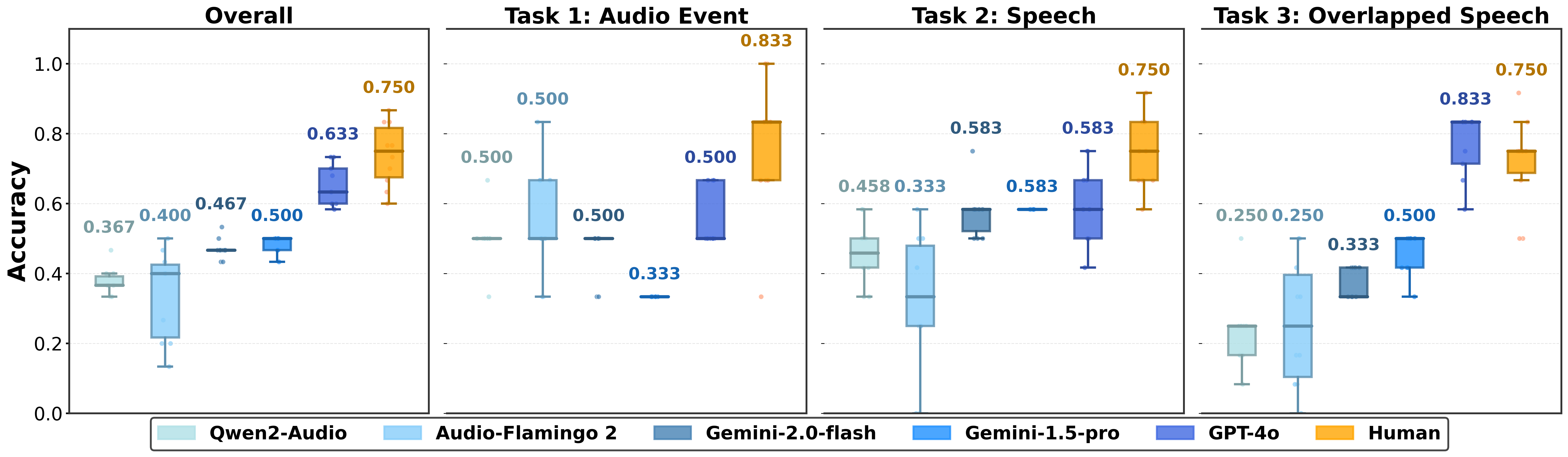}
    \caption{Performance of audio LLMs (without TTC) in comparison to human perception}
    \vspace{-10pt}
    \label{fig:allm}
\end{figure}

Figure~\ref{fig:allm} presents a comparative analysis of the performance of various audio LLMs and human perception in processing different auditory scenes. To simulate the variability in human auditory perception, we employed different temperature parameters within the same LLMs, allowing for a range of perceptual sensitivities.

Overall, all five tested audio LLMs underperform compared to human in processing auditory scenes of varying complexity. GPT-4o demonstrates the highest performance, followed by the Gemini model series, while Qwen2-Audio and Audio-Flamingo exhibit the lowest performance. This could potentially be due to the limited and less diverse data used for training the Qwen2-Audio models, and the smaller size (3B) of the Audio-Flamingo 2 model. This trend is also consistent in general across different auditory tasks, ranging from task 1 to task 3 with increasing levels of complexity.

\begin{wrapfigure}{l}{0.58\textwidth} 
    \centering
    \includegraphics[width=0.58\textwidth]{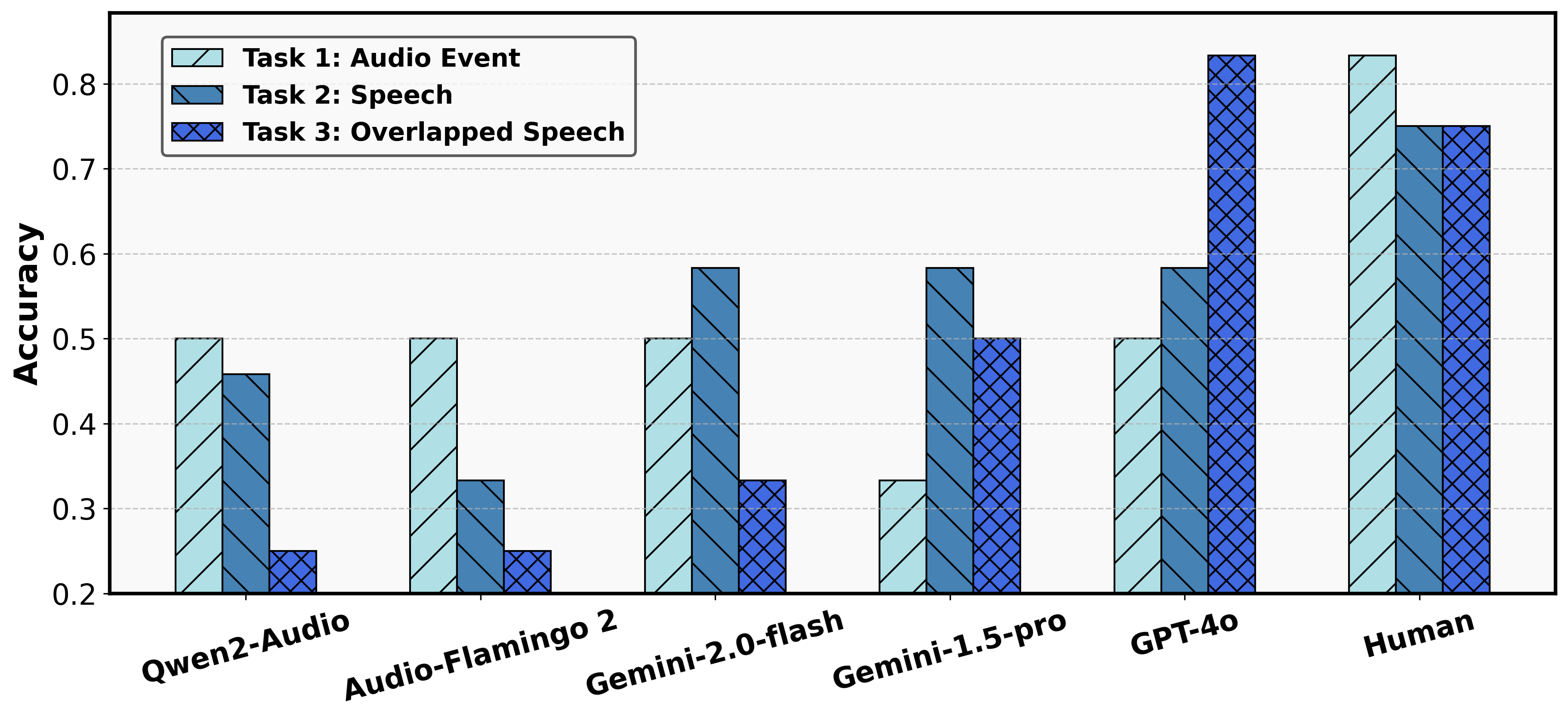} 
    \caption{Performance trend from simple to complex acoustic scenes.}
    \label{fig:allmtrend}
    \vspace{-10pt}
\end{wrapfigure}

Specifically, for Task 1 (audio event recognition) and Task 2 (speech comprehension), all audio LLMs perform similarly, with GPT-4o performing slightly better. Surprisingly, for task 3 of overlapping speech, GPT-4o performs better than the average human performance, although it does not surpass the best human performance. This suggests that the current GPT-4o can effectively process complex auditory conditions.

The trend across different auditory tasks is shown in Figure~\ref{fig:allmtrend}. A decline is observed as task complexity increases in general apart from GPT-4o and Gemini-1.5-pro. Qwen2-Audio, Audio-Flamingo 2 and Gemini-2.0-flash exhibit a performance decline similar to that of human perception as task difficulty increases. Interestingly, the performance of GPT-4o and Gemini-1.5-pro improves from simple audio event recognition tasks to more complex ones. This may be attributed to the broader coverage of speech training data compared to acoustic event data during the pre-training phase. 


\vspace{-5pt}
\subsection{Performance using TTC}
\begin{figure}[ht]
    \centering
    \includegraphics[width=1\linewidth]{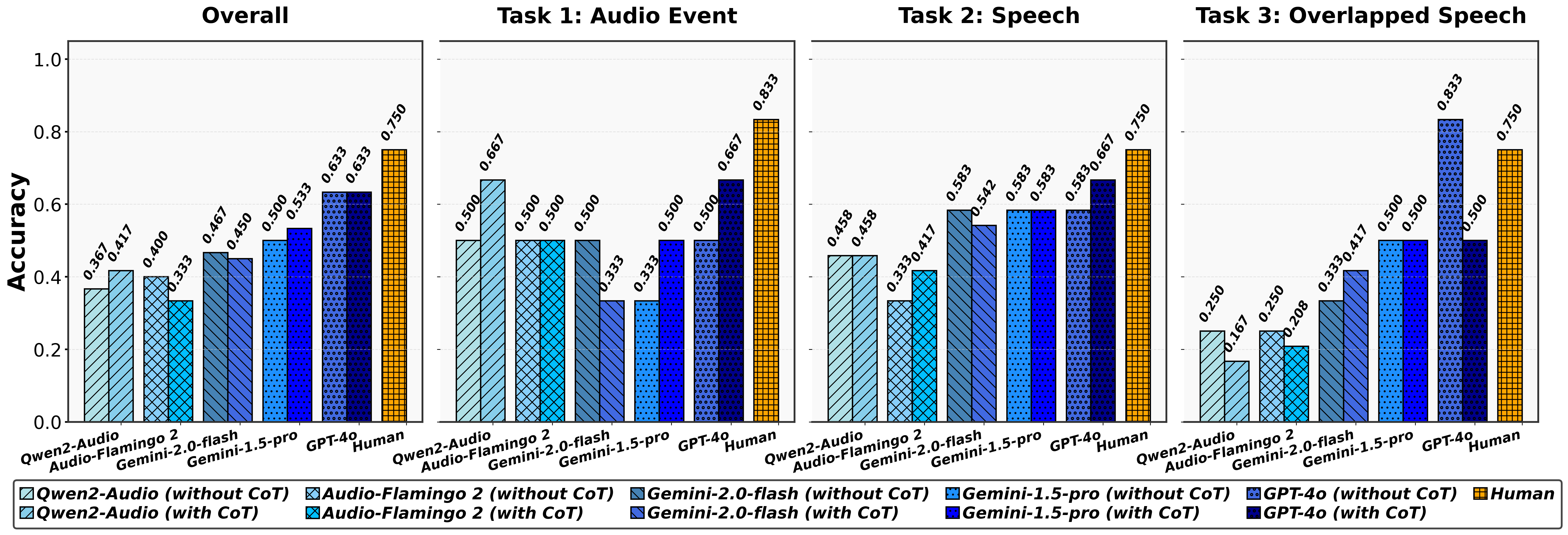}
    \caption{Evaluating the effect of CoT on model performance. The improvement with CoT varies depending on the specific models and tasks.}
    \vspace{-10pt}
    \label{fig:cot}
\end{figure}

\paragraph{CoT.} We initially applied CoT reasoning to enhance auditory inference capabilities. Figure~\ref{fig:cot} illustrates the performance improvements achieved through CoT compared to the original models without CoT prompting. The extent of these improvements varies depending on the specific models and tasks. Generally, CoT enhances the performance of Qwen2-Audio and Gemini-1.5-pro, while it does not yield significant benefits for other models. Notably, the improvements over the baseline without using TTC are primarily observed in task 2 of acoustic event recognition with 33.4\% and 50\% relative improvements for Qwen2-Audio and Gemini-1.5-pro, whereas more challenging auditory scenes do not exhibit similar gains. Interestingly, GPT-4o demonstrates a decline in task 3 performance when CoT is employed. These results suggest that GPT-4o may have been trained with reasoning capabilities, and therefore, CoT prompting does not further enhance its performance, or the CoT prompt may require careful design to be more effective. 

\paragraph{Search against verifiers.} Table~\ref{tab:search} presents the performance using various approaches of search against verifiers. The models Qwen2-Audio and Audio-Flamingo 2 were selected due to their open-source nature, which allows access to probabilities of outputs or search paths. Generally, all approaches improved performance compared to the baseline without using TTC. Specifically, for Qwen2-Audio, beam search yielded the best results, with significant improvements in Task 2 and Task 3, showing relative improvements of 63.8\% and 66.8\%, respectively. This indicates that exploring multiple decoding pathways can potentially uncover alternative answers when acoustic tasks are challenging, suggesting that considering other options can be beneficial.

\begin{table}
    \centering
    \caption{Performance comparison using different strategies of search against verifiers. Paraphrase reflects the corresponding improvements in percentage terms.}
    \definecolor{lightred}{rgb}{0.9,0.5,0.5}
    \definecolor{lightgreen}{rgb}{0.5,0.8,0.5}
    \large  
    \resizebox{1.01\columnwidth}{!}{%
    \begin{tabular}{l|c|c|c|c||c|c|c|c}
    \toprule
        & \multicolumn{4}{c||}{Qwen2-Audio} & \multicolumn{4}{c}{Audio-Flamingo 2} \\ \midrule
        & Overall & Task 1 & Task 2 & Task 3 & Overall & Task 1 & Task 2 & Task 3 \\
        \cmidrule{2-9}
        No TTC & 0.367 & 0.500 & 0.458 & 0.250 & 0.400 & 0.500 & 0.333 & 0.250 \\ \midrule
        CoT & 0.417 \textcolor{lightred}{(+13.6)} & \textbf{0.667} \textcolor{lightred}{(+33.4)} & 0.458 {(+0.00)} & 0.167 \textcolor{lightgreen}{(-33.2)} & 0.333 \textcolor{lightgreen}{(-16.8)} & 0.500 {(+0.00)} & 0.417 \textcolor{lightred}{(+25.20)} & 0.208 \textcolor{lightgreen}{(-16.8)} \\ \midrule
        Majority & 0.400 \textcolor{lightred}{(+9.00)} & 0.500 {(+0.00)} & 0.583 \textcolor{lightred}{(+27.3)} & 0.167 \textcolor{lightgreen}{(-33.2)} & 0.467 \textcolor{lightred}{(+16.8)} & 0.500 {(+0.00)} & 0.500 \textcolor{lightred}{(+50.20)} & 0.417 \textcolor{lightred}{(+66.8)} \\ 
        \midrule
        BS-W & \textbf{0.500} \textcolor{lightred}{(+36.2)} & 0.167 \textcolor{lightgreen}{(-66.6)} & \textbf{0.750} \textcolor{lightred}{(+63.8)} & \textbf{0.417} \textcolor{lightred}{(+66.8)} & 0.500 \textcolor{lightred}{(+25.0)} & 0.500 {(+0.00)} & 0.750 \textcolor{lightred}{(+125.2)} & 0.250 {(+0.00)} \\ \midrule
        LLM-Top 1 & 0.400 \textcolor{lightred}{(+9.00)} & \textbf{0.667} \textcolor{lightred}{(+33.4)} & 0.500 \textcolor{lightred}{(+9.2)} & 0.167 \textcolor{lightgreen}{(-33.2)} & \textbf{0.667} \textcolor{lightred}{(+66.8)} & 0.500 {(+0.00)} & \textbf{0.833} \textcolor{lightred}{(+150.2)} & \textbf{0.583} \textcolor{lightred}{(+133.2)} \\
        LLM-W & 0.400 \textcolor{lightred}{(+9.00)} & \textbf{0.667} \textcolor{lightred}{(+33.4)} & 0.500 \textcolor{lightred}{(+9.2)} & 0.167 \textcolor{lightgreen}{(-33.2)} & 0.633 \textcolor{lightred}{(+58.3)} & \textbf{0.667} \textcolor{lightred}{(+33.4)} & 0.667 \textcolor{lightred}{(+100.3)} & \textbf{0.583} \textcolor{lightred}{(+133.2)} \\
        \bottomrule
    \end{tabular}%
    }
    \label{tab:search}
    \vspace{-8pt}
\end{table}

For Audio-Flamingo 2, the most effective approach was the LLM verifier, indicating that advanced LLMs with a better understanding of acoustic scenes can enhance the performance of weaker LLMs in auditory tasks. Notably, the most significant improvements were observed in Task 2 and Task 3, particularly in Task 2 with 1.5 times improvements. This suggests that while these models may have some understanding of speech tasks, additional pathways could be advantageous when confronted with complex auditory scenes.

\begin{wrapfigure}{l}{0.5\textwidth} 
    \vspace{-10pt} 
    \centering
    \includegraphics[width=0.5\textwidth]{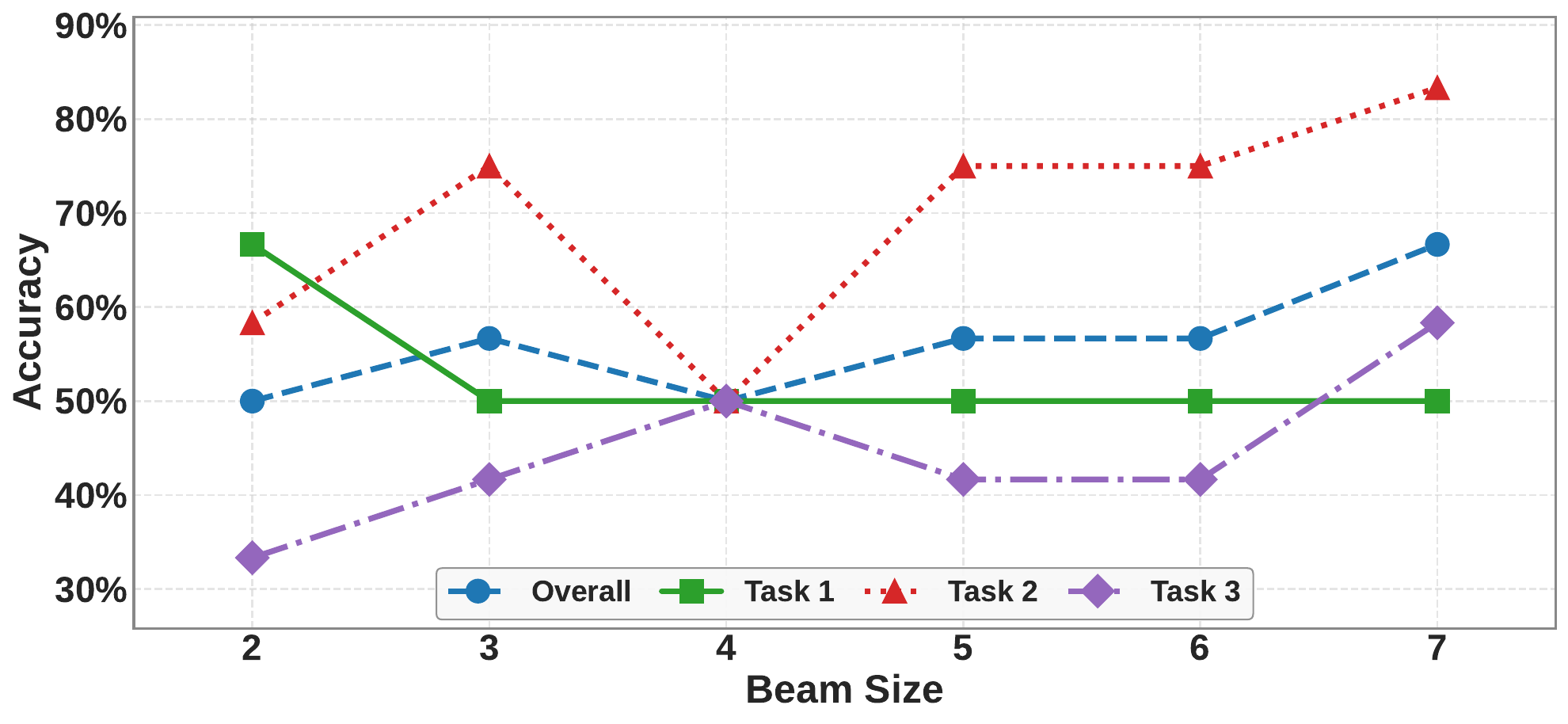} 
    \caption{Performance with different beam sizes using Audio-Flamingo 2 model.}
    \label{fig:bs}
\end{wrapfigure}

To gain a deeper understanding of how beam search contributes to the improved performance, we conducted a detailed performance analysis using different beam sizes with Audio-Flamingo 2, as illustrated in Figure~\ref{fig:bs}. The results indicate that as the beam size increases, overall accuracy improves, as well as for Task 2 and Task 3. This suggests that a larger beam size allows the model to generate multiple candidate answers, thereby increasing the likelihood of selecting a more accurate response. This is specifically beneficial when processing more challenging acoustic scenes, as the model's top candidate output may be less reliable. 
More ablation studies can be referred to Appendix~\ref{sec:abl}.

%% file: sections/06-conclusion.tex
\section{Conclusion}
This paper explored the inherent auditory cognitive capabilities of audio LLMs and proposed five test-time compute methods to enhance their performance across a range of complex auditory cognitive tasks. The results revealed the underperformance of current audio LLMs, particularly in handling complex auditory cognitive tasks. GPT-4 demonstrated the best performance and surprisingly surpassed some human perception capabilities in handling complex acoustic scenes. Additionally, the five test-time compute methods demonstrated significant improvements in enhancing auditory cognitive performance, paving the way for more effective audio LLMs in processing auditory tasks. This study offers a novel approach to improving audio LLMs during inference, and provides valuable insights for future research in audio LLMs for auditory cognition. Future studies will focus on expanding the dataset to evaluate more diverse auditory conditions and on developing the reward models specifically tailored for auditory tasks.

%% file: sections/07-appendix.tex
\newpage
\section*{Appendix}\label{appendix}
\appendix
\section{Methods}
\subsection{Prompt for Audio LLMs}\label{app:llmp}
The example prompts used for audio LLMs without CoT are shown in Table~\ref{tab:llmp}. For Gemini-series models and GPT-4o, the responses are clean with no explanation. For Qwen2-Audio and Audio-Flamingo 2, an additional constraint is added to ensure the output is clean as: "Just output the selected answer".

\begin{table}[h]
\centering
\caption{Example prompts for audio LLMs}
\begin{tabular}{|p{3.5cm}|p{9.6cm}|}
\toprule
\textbf{Task 1 Prompt}: & Focus on ANIMAL sound. Did you hear the cat meowing? Select the best option from the following: ['Yes', 'No']. No explanation is needed. \\ \midrule
\textbf{Task 2 Prompt}: & Listen carefully. What is the LAST TWO digits spoken by male? Select the best option from the following: ['6,2', '2,2']. No explanation is needed. \\ \midrule
\textbf{Task 3 Prompt}: & Focus on FEMALE. Which digit has NOT been mentioned by female? Select the best option from the following: ['4', '6', '8', '9']. No explanation is needed. \\
\bottomrule
\end{tabular}
\label{tab:llmp}
\end{table}

\subsection{CoT prompting}
In addition to the prompts used for audio LLMs, CoT prompting for three different tasks is additionally integrated, as shown in Table~\ref{tab:cot_tasks}.
\begin{table}[ht!]
\centering
\caption{CoT prompting for different tasks and acoustic scenes}
\begin{tabular}{|p{3.2cm}|p{9.8cm}|}
\toprule
 \textbf{Auditory Tasks} & \textbf{Chain of Thought (CoT) Prompting} \\ \midrule
\texttt{Task1: Audio Event} & There are different sound events in this audio trial. First, recognize each audio event, and then determine if the sound event mentioned in the question is present among these sound events. \\ \midrule
\texttt{Task2: Speech} & There is a sequence of digits spoken by either a male or a female speaker. First, recognize the gender and the digit for each spoken digit. Then, identify the gender of the speaker or speakers mentioned in the question, recall the corresponding part of the sequence, and find the digit or digits spoken by the identified gender or genders to answer the question. \\ \midrule
\texttt{Task3: Overlapped Speech} & There is a sequence of digits spoken simultaneously by both male and female speakers. First, identify the digits in the sequence spoken by the female and male speakers separately. Then, determine the gender of the speaker or speakers mentioned in the question. Recall the corresponding part of the sequence and find the digits spoken by the identified gender or genders to answer the question. \\ \bottomrule
\end{tabular}
\label{tab:cot_tasks}
\end{table}

\subsection{Prompt for LLM verifier}\label{app:llm}
The prompt for the LLM verifier is displayed in Table 4. To improve the reliability of the LLM verifier's scoring, we instructed the LLM backbones (i.e., Qwen2-Audio and Audio-Flamingo 2) to provide explanations for their decisions, offering more comprehensive information for scoring. These responses are further evaluated by the LLM verifier using the prompt in Table 4. 
\begin{longtable}{|p{3.5cm}|p{9.6cm}|}
    \caption{Prompt for LLM verifier} \\ 
    \toprule
    \rowcolor[gray]{0.85} \textbf{Section} & \textbf{Details} \\
    \midrule
\textbf{Audio Task Question} & \texttt{[Insert the audio task question here] \newline \textbf{Example}: "Focus on digits order spoken by FEMALE. What is the order of last three digits spoken by female?"} \\
    \midrule
    \textbf{Available Options} & \texttt{[Insert the answer options here] \newline \textbf{Example}: ["6,7,9", "8,9,1", "6,9,1", "8,7,9"]} \\
    \midrule
    \textbf{Machine's Response} & \texttt{[Insert the machine-generated response here] \newline \textbf{Example}: The sequence of digits spoken by the female is 6, 7, and 9. The sequence spoken by the male is 8, 9, and 1. Therefore, the last three digits spoken by the female are 9, 7, and 6. The answer is [6, 7, 9].} \\
    \midrule
    \textbf{Evaluation Process} & 
    \begin{minipage}[t]{\linewidth}
    \begin{itemize}
        \item Form your own understanding of the audio content first.
        \item Determine what you believe is the correct answer.
        \item Compare the machine's response to your independent assessment.
        \item Consider:
        \begin{itemize}
            \item \textbf{Accuracy:} Does the response align with what you heard in the audio?
            \item \textbf{Reasoning:} Does the explanation reflect genuine understanding?
            \item \textbf{Relevance:} Does it address what was actually asked?
        \end{itemize}
    \end{itemize}
    \end{minipage} \\
    \midrule
    \textbf{Scoring Criteria} & 
    \begin{minipage}[t]{\linewidth}
    \begin{itemize}
        \item A score closer to \textbf{0.0} indicates incorrect answers, poor reasoning, or irrelevant content.
        \item A score closer to \textbf{1.0} indicates correct answers with clear, accurate reasoning.
    \end{itemize}
    \end{minipage} \\
    \midrule
    \textbf{Evaluation Format} & 
    Please provide your evaluation in the following format: 
    \begin{itemize}
        \item RATING: [single decimal 0.0-1.0]
        \item ANALYSIS: [2-3 sentences explaining your rating]
        \item SELECTED OPTION: [Choose the best matching answer from the options above]
    \end{itemize} \\ \bottomrule
\end{longtable}

An example response from the LLM verifier is given in Table~\ref{tab:evaluation}.
\begin{table}[h]
\centering
\caption{An example response of LLM verifier.}
\begin{tabular}{|p{3.2cm}|p{9.8cm}|}
\toprule
\rowcolor{gray!50} 
\textbf{RATING}: & 0.8 \\
\midrule
\textbf{ANALYSIS}: & The machine's response correctly identifies the last three digits spoken by the female voice as 6, 7, and 9. However, there is a mistake in the explanation. The response states that the last three digits spoken by the female are 9, 7, and 6, which is incorrect. Despite this inconsistency, the final answer matches the audio content, and the noted mistake doesn't change the correctness of the identified digits. \\
\midrule
\textbf{YOUR ANSWER} & [6, 7, 9]
 \\
\bottomrule
\end{tabular}
\label{tab:evaluation}
\end{table}

\section{Experimental setup}
\subsection{Models}\label{app:setup}
The models used in this study includes five audio LLMs.
\begin{itemize}
\item Qwen2-Audio~\citep{chu2024qwen2}: Qwen2-Audio, developed by Alibaba, is a state-of-the-art multimodal model designed for processing and generating both text and audio. It is capable of handling various audio signal inputs to perform audio analysis or generate direct textual responses based on speech instructions. The model size utilized in our study is 7B parameters.

\item Audio-Flamingo 2~\citep{ghosh2025audio}: Audio-Flamingo 2, developed by NVIDIA, is an extension of the Flamingo model that enhances long-audio understanding and reasoning capabilities. It has demonstrated superior performance compared to other audio LLMs. Audio-Flamingo 2, with just a 3B parameter small language model, achieves state-of-the-art results across more than 20 benchmarks.

\item Gemini-2.0-Flash~\citep{google2023geminiflash}:  Gemini-2.0-Flash is a lightweight version of the Gemini-2.0 series developed by Google and updated in 2025, optimized for real-time audio processing with minimal latency. It maintains strong performance in speech recognition and synthesis while reducing computational requirements.

\item Gemini-1.5-Pro~\citep{team2024gemini}: Gemini-1.5-Pro is a more advanced version of the Gemini-1.5 series, featuring enhanced multimodal capabilities and improved contextual reasoning for audio-based tasks. It is trained on a larger dataset, enabling robust performance in speech-to-text conversion, emotion recognition, and multilingual audio processing.

\item GPT-4o~\citep{hurst2024gpt}: GPT-4o is OpenAI’s latest multimodal model, capable of processing and generating both text and audio with exceptional accuracy. It integrates a unified architecture for seamless cross-modal interactions, excelling in tasks such as audio transcription, voice synthesis, and conversational AI. 
\end{itemize}

\begin{figure}[ht]
    \centering
    \includegraphics[width=1.0\linewidth]{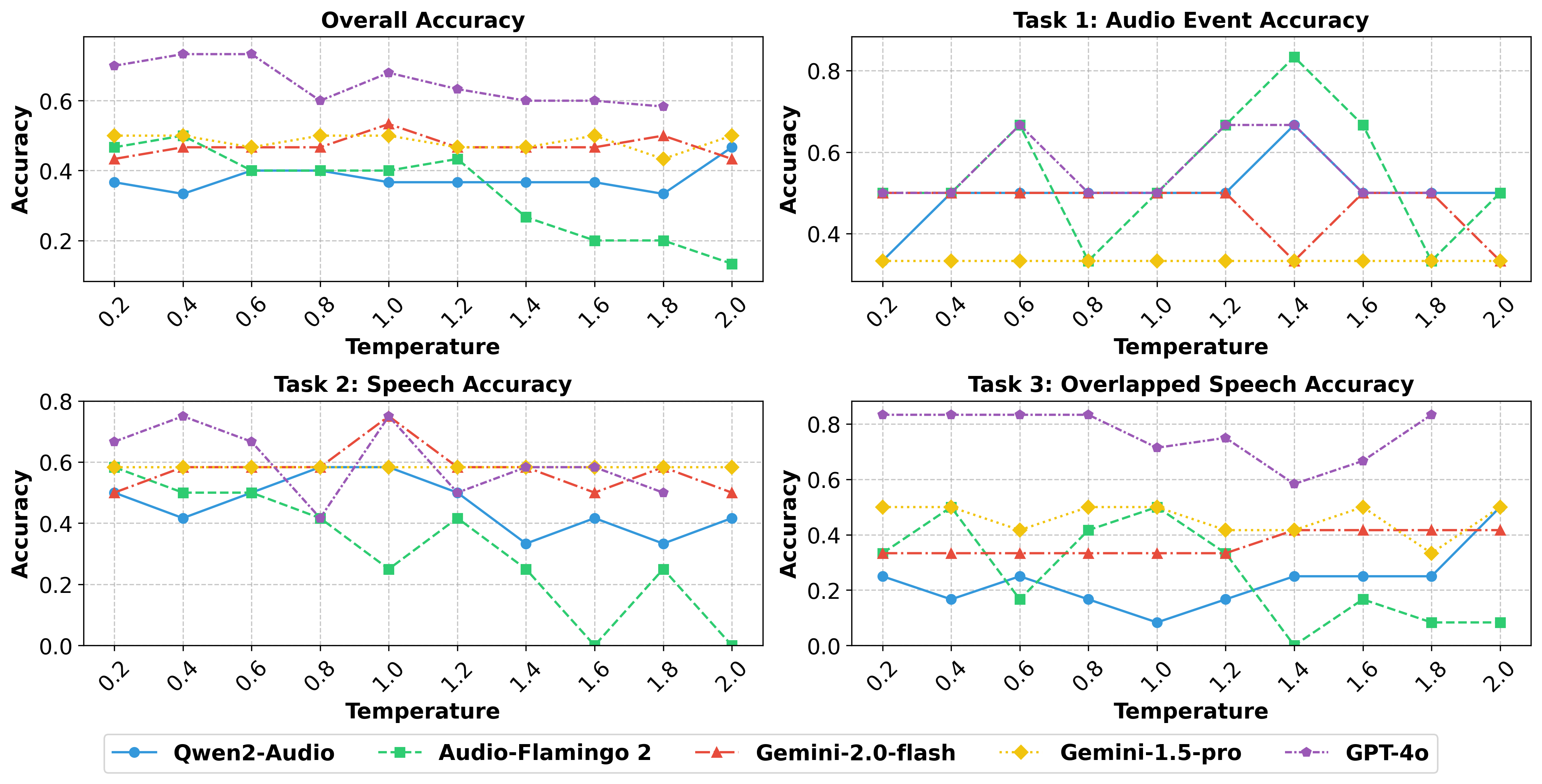}
    \caption{Impact of temperature}
    \label{fig:temprature}
\end{figure}

\subsection{Parameters}\label{app:para}
For the TTC majority vote baseline, we employed a range of temperature values from 0 to 2, with increments of 0.2. In terms of beam search, we utilized a beam count ranging from 2 to 7 constrained by memory limitations.

For the LLM verifier, GPT-4o was chosen as the verifier model based on our initial exploration of audio LLMs capabilities in audio comprehension tasks. This exploration demonstrated GPT-4o's superior performance in audio processing tasks compared to other audio LLMs. The verifier was prompted to score the generated \( N \) responses by considering both the audio inputs and the outputs from the first-stage LLM. 

\section{Ablation study}\label{sec:abl}
\paragraph{Impact of temperature}
Temperature controls the randomness of the model's output during text generation by influencing how the model samples from the probability distribution of possible next words. As shown in Figure~\ref{fig:temprature}, different models exhibit varying patterns in response to different temperature settings. A temperature of 2.0 was not evaluated for GPT-4o, as it generates nonsensical responses using temperature of 2.0. Additionally, the temperature shows distinct patterns across different tasks, suggesting that the model may require different temperature settings tailored to optimize performance for various acoustic scenes.